\documentclass{ws-procs975x65}

\allowdisplaybreaks

\begin{document}
\title{Circular inspiral templates}

\author{MANUEL TESSMER}
\address{Theoretisch-Physikalisches Institut, Friedrich-Schiller-Universit\"at,\\
Max-Wien-Platz\ 1, 07743 Jena, Germany, EU}

\begin{abstract}
Recently, a new class of restricted gravitational wave search templates, termed the
TaylorEt template was proposed for the search of inspiralling compact
binaries.
The TaylorEt approximant is different from the usual time-domain post-Newtonian
approximants in that it employs the orbital binding energy rather than
the orbital frequency or the closely related parameter ``x''.
We perform detailed studies to probe the fitting factors of
TaylorEt at 3.5pN for nonspinning comparable mass compact binaries
vis-a-vis the TaylorT1, TaylorT4, and TaylorF2 at 3.5pN approximants
in LIGO,  Advanced LIGO  and Virgo interferometers.
\end{abstract}

\bodymatter

\section{Templates and LAL routines}
In this article, we consider data analysis performances of several inspiral
templates for nonspinning compact binaries, used by the LSC Algorithms Library (LAL).
These routines are compared with the TaylorEt template with respect to
the numerical values of the fitting factor (FF) and the faithfulness
[\citen{Bose:Gopakumar:Tessmer:2008}].
The task is to find out whether they can be used as effective and faithful search
templates if the detectors strain data is assumed to provide the
TaylorEt approximant. We give a short introduction to all the considered
post-Newtonian (pN) model waveforms.

\begin{itemize}
 \item
The TaylorT1 approximant [\citen{Damour:Iyer:Sathyaprakash:2001}]
employs the instantaneous gravitational wave (GW) angular velocity $\omega$ and  $\mathcal L$, 
the far-zone flux 
of the pN accurate energy $E$, 
for an evolution equation of both GW phase $\phi$ and $\omega$ at time $t$,
\begin{align}
h(t) & \propto x\, \cos (2\,\phi(t)) \,, 													\label{Eq::TaylorT1}\\
\frac{{\rm d} \phi (t)}{{\rm d}t} & = \omega (t)  \equiv \frac{c^3}{G\,m}\, x^{3/2}\,,			\label{EqP1b}\\
\frac{{\rm d} x(t)}{{\rm d} t} &=  -\frac{{\cal L}( x)}{ \left( {\rm d} {E} / {\rm d}  x \right)}\,. 	\label{EqP1c}
\end{align}

\item
The TaylorT4 approximant (not used in LAL) is a slightly modified version of T1 and uses, as the only difference to T1, a
Taylor-expanded version of the ratio in Eq. (\ref{EqP1c}).

\item
The TaylorF2 approximant is the Fourier domain pendant to TaylorT1, obtained by a stationary phase approximation
[\citen{Damour:Iyer:Sathyaprakash:2000, Arun:Iyer:Sathyaprakash:Sundararajan:2005}].

\item
The TaylorEt approximant  [\citen{Gopakumar:2008}] uses the orbital binding energy as the quantity to be evolved by radiation reaction,
\begin{eqnarray}
\label{Eq::TaylorEt}
 \frac{{\rm d} \phi}	{{\rm d} t}	&=& E^{3/2} \, \left\{1 +  \,\dots\,  + E^3\,[\dots] \right\}, \\
 \frac{{\rm d} E}	{{\rm d}  t}	&=& \frac{64}{5}\, \eta \, E^{5} \, \left\{1 +  \,\dots\,  +E^{7/2}[\dots] \right\},
\end{eqnarray}
where $\eta$ is the symmetric mass ratio $\eta \equiv m_1 \, m_2 /{(m_1+m_2)}^2$.
\end{itemize}
For our studies, the post-Newtonian expansions of Eq. (\ref{EqP1c}),
GW luminosity $\mathcal {L}$, and expressions in Eq. (\ref{Eq::TaylorEt})
are used up to 3.5pN order. All the necessary references and explicit formulas,
partly only given with dots here, can be extracted from the investigation we like to summarize shortly
[\citen{Bose:Gopakumar:Tessmer:2008}].

\section{Issues in GW phasing and data analysis}
For our data analysis considerations, the {\em time domain} GW forms are constructed
in the following way. The orbital evolution is set to start at the initial instant $t_0$ when the
instantaneous GW frequency of $f_{GW}$ kicks in the detectors bandwidth. It is necessary to
transform $f_{GW}$ into the binding energy in a pN accurate way for the Et waveform and
to let $f_{GW}$ or $E$ evolve due to the model. The upper boundary for the GW frequency
is that of the last stable circular orbit of a black hole having the total mass $m$ of the binary
system.
The merger and ringdown phases are excluded from our analysis.

\section{Results and summary}
For the equal-mass case, the investigation showed up that for all three employed interferometer models,
the systematic errors of the estimated masses continuously rise from $0$ to the order of
magnitude of $\sim 6-8\%$ when going from $m=1.4$ to $m=40$ solar masses,
supposing that the TaylorEt is the signal and TaylorT1 and T4 are taken to be the templates.
The fitting factors have a minimum at the mass in the signal of $\sim 10 \, M_\odot$, which is always
below $96\%$. Additionally, the FF values are {\em always} below $96\%$ for advanced LIGO
and Virgo for all the considered cases.
For the unequal-mass case, the results behave qualitatively somewhat different. To give representative numbers,
let us show the following table
for $\eta=0.1875$ (for the $m_1/m_2=1/3$ signal family) and advanced LIGO at 3.5pN order, presenting
the FFs, the ratios of the estimated total mass with respect to the signal's one, and the estimated $\eta$,
as functions of template masses in the first row.

\begin{table}[!h]
\begin{tabular}{|c|c|c|c|c|c|}
\hline ~
$\frac{m_1}{M_\odot}$ : $\frac{m_2}{M_\odot}$  & $m$ ($M_\odot$)   & template ~&~ FF ~&~ $m_{\rm est}/ M_\odot$~&~ $\eta_{\rm est}$~ \hfill \\
\hline
3:9	~&~ 12	~&~ T1  ~&~ 0.98	~&~ 10.16 ~&~ 0.250 ~\\
        ~&~     ~&~ F2  ~&~ 0.97 ~&~ 10.23 ~&~ 0.248 ~\\
\hline
5:15	~&~ 20	~&~ T1  ~&~ 0.96	~&~ 16.91 ~&~ 0.250 ~\\
        ~&~     ~&~ F2  ~&~ 0.97 ~&~ 16.96 ~&~ 0.250 ~\\
\hline 
10:30	~&~ 40	~&~ T1  ~&~ 0.98	~&~ 33.60 ~&~ 0.250 ~\\	
        ~&~     ~&~ F2  ~&~ 0.95 ~&~ 33.82 ~&~ 0.249 ~\\
\hline
\hline
\end{tabular}
\caption{FFs \& estimated ($m \,, \eta$) for several signal masses, $\eta=0.1875$}
\label{tab:TableFFsAdLIGO}
\end{table}
As it is clearly visible, our code evaluated the FF at large biases of the estimated total mass $m_\text{est}$
and mass ratio $\eta_\text{est}$.
We cannot give a full overview here and like to refer to the more detailed presentation of the numerical results
[\citen{Bose:Gopakumar:Tessmer:2008}].

One result of this investigation is that fitting factors for $\ge 95$\% for unequal-mass
binaries can be only achieved at the cost of highly biased $m_\text{est}$ and $\eta_\text{est}$.
The faithfulness for the T1 and Et comparison turned out to be $\sim 0.45$.

This investigation is only able to sketch the efficiency of capturing one model with the
other. The question to be asked is what kind of analytic waveform is closer to reality
can partly be answered by numerical relativity (NR), which is only able to model few GW cycles
at the end of the binary inspiral. It turned out that the TaylorT4 waveform is, for some
cases, very close to NR.
With respect to other waveforms that are closer to NR in the sense of matched filtering,
it has unfavorable properties like artificial introduction of large parameter biases.
This investigation confirms what has been presented  in more detail in another recent work
[\citen{Buonanno:Iyer:Ochsner:Pan:Sathyaprakash:2009}]
and has been extended for TaylerEt to eccentric orbits 
[\citen{Tessmer:Gopakumar:2008:2}], where in the latter case we are left
with the demanding question for numerical results for the eccentric inspiral.

\section*{Acknowledgements}

I wish to thank A. Gopakumar and S. Bose for helpful discussions.
Special thanks go to G. Sch\"afer for encouragement.
This work is supported  by
SFB/TR 7 ``Gravitational Wave Astronomy'' of DFG
and through ``LISA Germany''
of DLR (Deutsches Zentrum f\"ur Luft- und Raumfahrt).

\bibliographystyle{utphys}



\end{document}